# A Compact LSTM-SVM Fusion Model for Long-Duration Cardiovascular Diseases Detection

1st Siyang Wu
*college of LSA*
*University of Michigan*
MI, USA
siyangwu@umich.edu

*Abstract*— Globally, cardiovascular diseases (CVDs) are the leading cause of mortality, accounting for an estimated 17.9 million deaths annually. One critical clinical objective is the early detection of CVDs using electrocardiogram (ECG) data, an area that has received significant attention from the research community. Recent advancements based on machine learning and deep learning have achieved great progress in this domain. However, existing methodologies exhibit inherent limitations, including inappropriate model evaluations and instances of data leakage. In this study, we present a streamlined workflow paradigm for preprocessing ECG signals into consistent 10-second durations, eliminating the need for manual feature extraction/beat detection. We also propose a hybrid model of Long Short-Term Memory (LSTM) with Support Vector Machine (SVM) for abnormality detection. This architecture consists of two LSTM layers and an SVM classifier, which achieves a SOTA results with an Average precision score of 0.9402 on the MIT-BIH arrhythmia dataset and 0.9563 on the MIT-BIH atrial fibrillation dataset. Based on the results, we believe our method can significantly benefit the early detection and management of CVDs.

## I. INTRODUCTION

Cardiovascular diseases (CVDs) have become the most prevalent health condition globally, accounting for over 17.3 million deaths, with a projected increase to over 23.6 million by 2030 [1]. In general, the mortality rate of CVDs is twice that of cancer annually [2]. In an analysis across 197 countries, the direct and indirect economic cost of heart failure reached up to $108 billion in 2012 [3]. However, the majority of cases could be prevented through early detection and hospitalization.

Cardiac arrhythmia, particularly atrial fibrillation (AFIB), is strongly linked to severe CVDs such as stroke and congestive heart failure [4, 5]. Traditionally, cardiac arrhythmia is diagnosed by analyzing electrocardiogram (ECG) data collected by a Holter monitor, which is a 12-lead ECG system used for 24-hour monitoring and can be burdensome for patients [6]. With advancements in technology, wearable devices are now capable of recording long-term single-lead ECG data. While it is true that the clinical use of single-lead ECG has limitations compared to 12-lead ECG data [7], this does not negate the potential for crucial information to be obtained through long-term surveillance using single-lead ECG data collected from wearable devices [8, 9].

In this paper, we are going to focus on a specific field of ECG analysis which relies on wearable devices for ECG data collection and AI for ECG data analysis (ECG-AI) [10] and in ECG-AI we are going to talk about the work based on single lead ECG analysis. There are extensive explorations in this field and their works end up having satisfactory performance in distinguishing abnormalities in ECG data empowered by deep learning and machine learning techniques. A Deep Genetic Ensemble of Classifiers is proposed by Plawiak et al. for arrhythmia detection [11]. Ribeiro et al. proposed a CNN deep neural network method for arrhythmia detection [12]. A CNN-Bidirectional Long Short-Term Memory was raised by Shin et al [13]. to perform general arrhythmia detection. There are also attempts to detect AFIB and have seemingly good results. Deep learning Neural Network was implemented to perform the task of AFIB detection by Cheng et al. [14]. A Cat-boost Decision Tree Ensemble method proposed by Shao et al [15] to detect AFIB. Although their methodologies made great progress and fresh ideas in the field of ECG-AI, some detrimental flaws are hidden in their methodologies which makes their results dubious. Each of them adopts an intra-patient paradigm which is considered to be inappropriate and problematic to be generalized to clinical utilization [16-18]. Meanwhile, they neglect the fact that MIT BIH Atrial Fibrillation Dataset and MIT BIH Arrhythmia Dataset are imbalanced in terms of abnormal vs normal which leads to a problem that the evaluation metrics that they stick to are usually not representative of the true performance of their model [20, 21].

In this paper, we aim to address the aforementioned problems by adhering strictly to an inter-patient paradigm and utilizing the average precision score (AP score) which is equivalent to of Area Under the Curve Precision and Recall (AUC PR) to evaluate model performance. AUC PR is established as a more reliable method of assessing the performance of models in binary classification tasks involving imbalanced datasets, as demonstrated in [20]. In addition, we propose a new model called Long Short-Term Memory and Support Vector Machine Fusion method (LSF), specifically designed for processing time series data, and performing classification tasks. To better evaluate the efficacy of the LSF model, we trained a baseline model consisting solely of LSTM layers that are the same size as those in our hybrid model. We tested our models on the MIT BIH arrhythmia database [22] and the MIT BIH Atrial Fibrillation database [23] and established baseline AP scores of 0.9235 and 0.9213, respectively.

The contributions of this paper are as follows:

- Design and implement an LSF model which improved the AP score from 0.9235 to 0.9402 on the MIT BIH

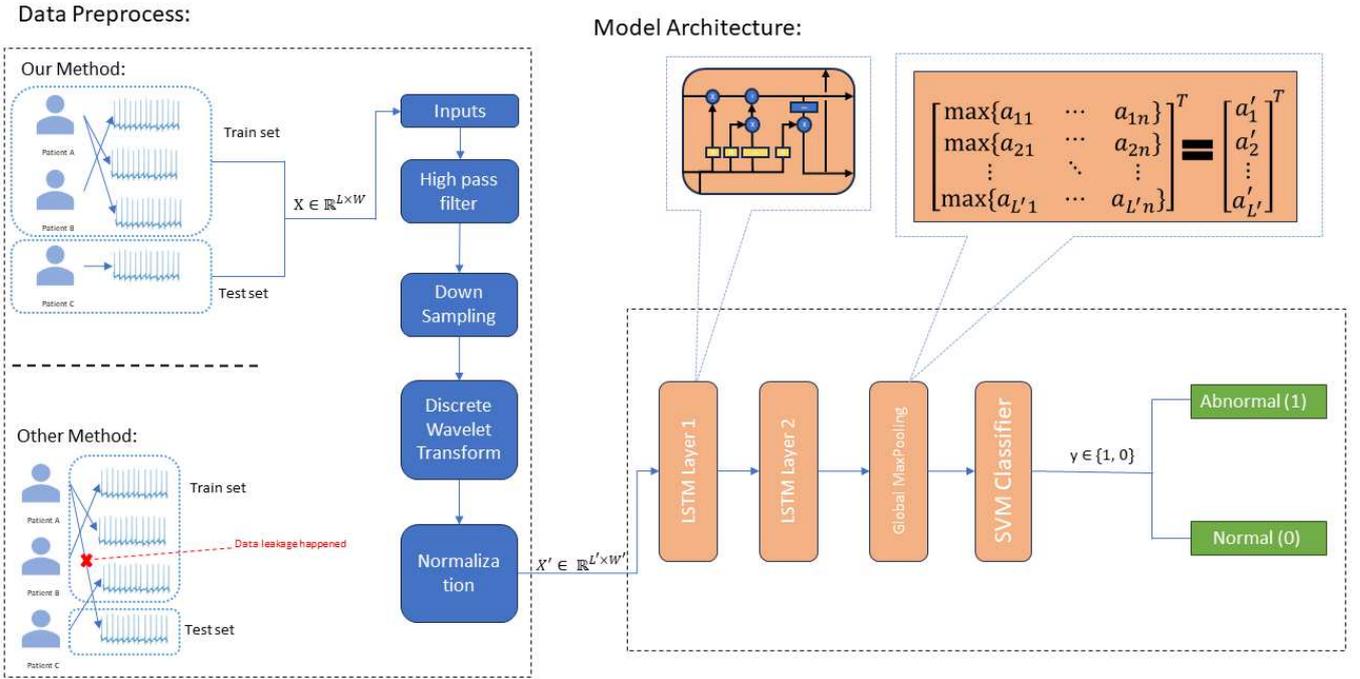

Fig 1, Workflow of LSF method

Arrhythmia dataset, and from 0.9213 to 0.9563 on the MIT BIH Atrial Fibrillation dataset.

- The time required for a 10-second segment on average was 0.0103 seconds, which is 31% faster than the method outlined in Yildirim et al. [19] that requires 0.015 seconds for a segment of the same length.
- Our proposed model provides a comprehensive, end-to-end workflow that enables a seamless transition from the raw signal input to the classification output without the necessity of manual feature extraction or QRS complex detection.

## II. RELATED WORK

In the field of ECG-AI, there have been various studies on detecting abnormalities from ECG data. Several methods focus on analyzing the ECG data at the heartbeat level. In the method proposed by Zhang et al. [24], a local window is created containing 300 samples around the R peak (99 samples before and 200 samples after) using the R peak annotation inherent in the MIT BIH Arrhythmia Dataset. Yu et al. [25] also use a similar beat-level segmentation method, creating a window containing 433 samples around the R peak. Chen et al. developed an RR Interval Extraction Based on QRS Waves to generate beat-level segmentations from ECG data for deep learning [26]. Although their methods yield satisfactory performance, the high reliance on an accurate R peak detection algorithm restricts the flexibility of these approaches. This issue can be addressed by segmenting the ECG signal into longer durations, encompassing multiple heartbeats. Instead of relying solely on the R peak detection method, we can train neural networks to select features that assist in the detection of abnormalities throughout the training process.

The model architectures vary from one another. Plawiak et al. [11] propose a Deep Genetic Ensemble of Classifiers (DGEC) for ECG abnormalities detection. The model consists of three major layers: the first layer comprises 48 classifiers, the second layer has 4 classifiers, and the third layer consists of a Support Vector Machine judge classifier. Yildirim et al. [19] develop a 16-layer 1D Convolutional model. Mazumder et al. propose a 14-layer model that combines Convolutional Neural Network and LSTM Neural Network. These methods bring fresh ideas to the field of ECG-AI and perform well. However, their models tend to be unnecessarily complex, making the methods more complicated. We believe that a simpler model can effectively address the problem of abnormalities detection from ECG. Therefore, we have designed and developed a 5-layer model comprising mainly 2 LSTM layers and 1 SVM judge classifier.

It is worth noting that there are controversies between the inter-patient paradigm and the intra-patient paradigm in the ECG-AI field. The intra-patient paradigm refers to a data separation method where, after segmentation and labeling, the train/test sets are randomly sampled from the segmentation pool. On the other hand, the inter-patient paradigm involves assigning patients to the train/test sets before any sampling is applied [16]. The intra-patient paradigm is widely used, including in the methods mentioned in the previous paragraphs [11-15, 24-27]. However, it has been proven to have problems with generalization for clinical use and the risk of data leakage. When using the intra-patient paradigm, particularly for ECG abnormalities detection, models are often tested on data with rhythms from the same patients they were trained on, which increases the likelihood of overfitting and bias [16-18]. In a study conducted by Minh et al. [18], the scores associated with the same model were found to be lower when using the more robust and rigorous inter-patient paradigm compared to the intra-patient paradigm. To address this issue, we will strictly follow the inter-patient paradigm by splitting patients into distinct train/test datasets to prevent data leakage.

## III. PROPOSED APPROACH

In order to address the issues raised in previous paragraphs. An Long short term memory and Support Vector

Machine Fusion (LSF) method is proposed in this paper and the pipeline of such method can be decompose into the following two parts:

$$X' = \Phi(X) \quad (1)$$

$$\hat{y} = f(X') \quad (2)$$

Figure 1 illustrates the workflow of this method. The input $X \in \mathbb{R}^{L \times W}$, which represents the physical signal of a single lead ECG data of length L and width W (W = 1 is used in this paper), undergoes data preprocessing function $\Phi(\cdot)$ such as high-pass filtering, down-sampling, and Discrete Wavelet Transform, resulting $X' \in \mathbb{R}^{L' \times W'}$, universally $L' < L$ and $W' = 2$. Then the processed data $X'$ pass through the model architecture $f(\cdot)$ which comprises two LSTM layers and an SVM classifier. This model architecture produces a binary value $\hat{y} \in \{1, 0\}$, where 1 indicates abnormal and 0 indicates normal..

A. Loss functions.

Abnormality detection is essentially a binary classification task. Our proposed method LSF consists of two parts: an LSTM block and an SVM classifier. The details of SVM can be found in section III, C, (3) since the cross entropy loss function is not applicable. The optimization process of the LSTM blocks ultimately uses the binary cross-entropy loss function defined as follows:

$$L(y, \hat{y}) = -\frac{1}{N} \cdot \sum_{i=0}^{N-1} y^{(i)} \cdot log(\hat{y}^{(i)}) + (1 - y^{(i)}) \cdot log(1 - \hat{y}^{(i)}) \quad (3)$$

where $\hat{y}^{(i)}$ is the predicted probability of label of $i$th item in the set, and $y^{(i)}$ is the true label.

B. Data Acquisition

Since 1975, the Massachusetts Institute of Technology, in collaboration with Boston's Beth Israel Hospital (now the Beth Israel Deaconess Medical Center), has been collecting and processing ECG data from patients. These datasets are typically hosted and served by PhysioNet [28]. In this paper, we will be working with two specific datasets: the MIT-BIH Arrhythmia Dataset [22] and the MIT-BIH Atrial Fibrillation Dataset [23].

C. Preprocess

The preprocess step consists of two modules, filter, and sample, as shown in Figure 2. The purpose is to transfer raw signal to feature vector for model.

1) High pass filter

Firstly, a high-pass filter is applied to the raw signal in order to eliminate baseline wander caused by the patient's movement or breath. This filter reduces the amplitude of the signal below a designated cutoff frequency $f_c$. The filter can be characterized by a transfer function $\Psi(\cdot)$ with coefficients (b, a), that are determined by the cutoff frequency $f_c$ sample frequency $f_s$, and order of the transfer function D. The transfer function is shown as follows:

$$\Psi(X_t) = \frac{\sum_{j=0}^{D} b_j X_t^{D-j}}{\sum_{j=0}^{D} a_j X_t^{D-j}} \quad (4)$$

Where $X_i$ is the value at $t^{th}$ place in raw single X.

2) Down sample

The raw signals in the MIT-BIH Arrhythmia and MIT-BIH Atrial Fibrillation datasets are sampled at rates of 360Hz and 250Hz, respectively. However, these high sampling rates often capture duplicate information and can slow down our computational speed, especially when working with LSTM blocks in deep learning. To overcome this issue, a polynomial resampling method is employed in this paper to uniformly downsample the signals from 360Hz and 250Hz to 100Hz. This downsampling technique effectively reduces data redundancy without significant loss of information.

3) Discrete Wavelet transforms

The Discrete Wavelet Transformation (DWT) is a technique that can decompose the original signal into high frequency and low frequency domains, while preserving the time series information. This is particularly beneficial for analyzing ECG data, where time is a crucial factor. The DWT is also computationally efficient compared to the Continuous Wavelet Transformation (CWT), which analyzes the signal at every possible point.

In this paper, the Haar wavelet is selected for performing a decomposition at stage one. The algorithm for the 1D Haar DWT involves applying two Haar filters: a low pass filter h[k] and a high pass filter g[k]. These filters, with a size of M = 2, are presented below:

$$h[k] = \left[\frac{1}{\sqrt{2}}, \frac{1}{\sqrt{2}}\right] \quad (5)$$

$$g[k] = \left[\frac{1}{\sqrt{2}}, -\frac{1}{\sqrt{2}}\right] \quad (6)$$

Given a 1-D signal S with size L, the approximation coefficient (cA) containing the information in low frequency domain and detail coefficient (cD) containing the information in high frequency domain can be acquired by the following two operations of convolution respectively:

$$cA[n] = \sum_{k=0}^{M-1} S[n-k] \cdot h[k], for\ n = 0, 1, .. L + M - 2 \quad (7)$$

$$cD[n] = \sum_{k=0}^{M-1} S[n-k] \cdot g[k], for\ n = 0, 1, .... L + M - 2 \quad (8)$$

Note that cA and cD are of the same size as signal S. However, a downsampling method is applied by keeping only the even terms to remove redundant information. Therefore, the cA and cD at stage 1 can be shown as follows:

$$cA_1 = downsample(cA) \quad (9)$$

$$cD_1 = downsample(cD) \quad (10)$$

4) Normalization

A Z-score normalization is applied as the final step before the data is fed into LSF. To further prevent data leakage, only

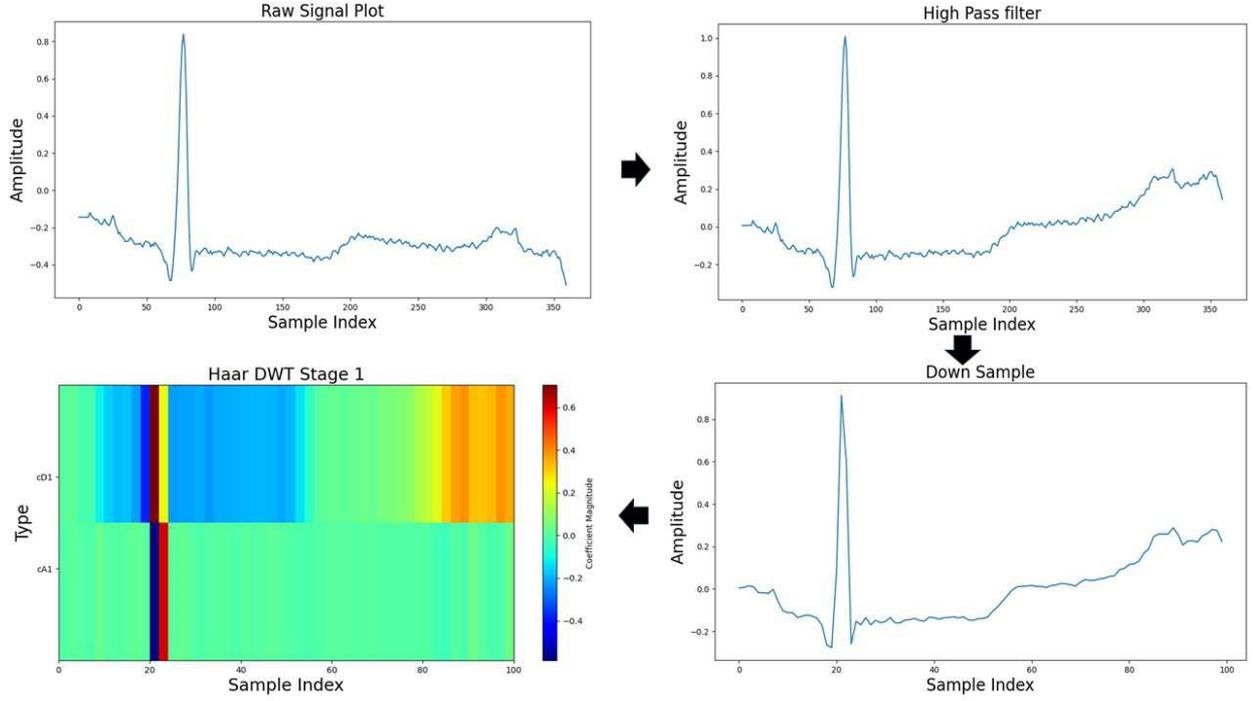

*Fig2. Data Preprocess Visualization*

the mean $\mu_{train}$, and $\sigma_{train}$ from the training dataset is used to normalize the train, test, future incoming dataset. The method of normalization is shown below:

$$Z = \frac{X' - \mu_{train}}{\sigma_{train}} \quad (11)$$

Where the X' is a feature vector after previous preprocessing pipeline, meanwhile $\mu_{train}$, and $\sigma_{train}$ are of the same size as X'. The operation in (11) is element-wise operation of a multi-dimensional vector.

C. Model Architecture

The LSF method mainly consists of two LSTM layers and an SVM classifier, as shown in Figure 1. The details are introduced as follows.

1) Long Short Term Memory Block

The LSTM (Long Short Term Memory) is a variation of the Recurrent Neural Network (RNN) designed to mitigate the vanishing gradient problem inherent in traditional RNNs that can result in the loss of information over long-range dependencies in a sequence of data. The LSTM introduces a long-term memory mechanism that enables it to capture important information across the entire sequence, making it well-suited for processing ECG data and other time series data, especially those with long durations, as in the segmentation method we employed.

Inside a LSTM module, at a certain timestamp $t$, there are an input gate $i_t$, a forget gate $f_t$, an output gate $o_t$, as well as a memory cell $c_t$, There components are responsible for reading, writing, and updating the information that the current cell should retain, as well as determining what should be passed to the memory cell at the next timestamp. The input gate, forget gate, output gate and activation of memory cell $\tilde{c}_t$ at current time stamp $t$ can be calculated based on the input data $X'_t$ and hidden state at last timestamp $h_{t-1}$:

$$i_t = sigmoid(U_i X'_t + V_i h_{t-1} + b_i) \quad (12)$$
$$f_t = sigmoid(U_f X'_t + V_f h_{t-1} + b_f) \quad (13)$$
$$o_t = sigmoid(U_o X'_t + V_o h_{t-1} + b_o) \quad (14)$$
$$\tilde{c}_t = tanh(U_c X'_t + V_c h_{t-1} + b_c) \quad (15)$$

The memory cell at current state can be updated by the following operation:

$$c_t = f_t \odot c_{t-1} + i_t \odot \tilde{c}_t \quad (16)$$

The hidden state $h_t$ can be computed:

$$h_t = o_t \odot tanh(c_t) \quad (17)$$

Note that the $\odot$ is the Hadamard product (element wise product of two vector of same size). $h_t \in \mathbb{R}^u$ where u is the number of units of each cell. This process repeats for $L'$ times where $L'$ is the total number of time stamp that feed into the LSTM block.

After the first block of LSTM process the data, the hidden state matrix $H^{1st} = [h_1^{1st}, h_2^{1st} \cdots h_{L'}^{1st}], H_1 \in \mathbb{R}^{L' \times u}$ formed by hidden state of each timestamp is feed to another LSTM block for a next round of processing and output $H^{2nd}$.

2) Global MaxPooling

Under the context of LSF, Global MaxPooling refers to the process of selecting the most prominent data point at a specific position within a hidden state over multiple timestamp. This process serves the purpose of reducing dimensionality and mitigating overfit. The exact process can be shown as follows:

$$GlobalMaxPooling(H^{1st}) = \begin{bmatrix} \max\{h_{11}^{1st} & \cdots & h_{1L'}^{1st}\} \\ \max\{h_{21}^{1st} & \cdots & h_{2L'}^{1st}\} \\ \vdots & \ddots & \vdots \\ \max\{h_{u1}^{1st} & \cdots & h_{uL'}^{1st}\} \end{bmatrix}^T \quad (18)$$

The output of $GlobalMaxPooling(H^{1st})$ is a feature vector $v \in \mathbb{R}^u$ which is believed to contain information of ECG data compressed in the most meaningful and efficient way.

3) SVM classifier

In the regular approach, the processed vector v obtained from the previous step is typically fed into a fully connected layer with a sigmoid activation function. This layer generates

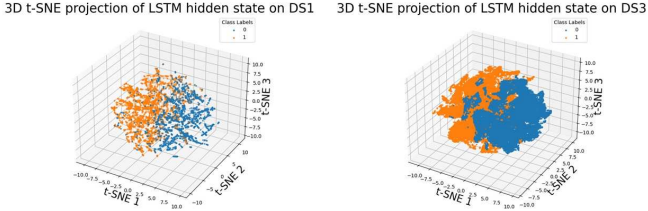

Fig3. 3D t-SNE projection of hidden state from LSTM

the probability of the attribution of the input signal X'. To gain insights into the feature vector v, we visualized the training set data using t-distribution stochastic neighbor embedding (t-SNE) [29]. This technique reduces the high-dimensional data into 3 dimensions for better visualization. Upon visualizing the feature vector v, which has a size of 100 array, we observed that the data clusters are separated into 2 distinct groups based on their labels. This can be seen in Figure 3. This distribution is especially suitable for SVM to perform binary classification on.

The feature vector $v$ is feed into soft margin SVM. The primal objective function can be shown as:

$$\min_{w,b,\xi} \frac{1}{2} w^T w + C \sum_{i=1}^{n} \xi_i \quad (19)$$

$$s.t.\ y^{(i)}(w^T + b) \geq 1 - \xi_i \quad for\ \xi_i \geq 0\ and\ \forall i \in \{1,\dots,n\} \quad (20)$$

In order to improve computation efficiency and add nonlinearity the primal form will then be converted into to the Lagrangian dual form shown as:

$$\max_{\alpha} \sum_{i=1}^{n} \alpha_i - \frac{1}{2} \sum_{i=1}^{n} \sum_{j=1}^{n} \alpha_i \alpha_j y^i y^j K(v^i, v^j) \quad (21)$$

$$s.t.\ 0 \leq \alpha_i \leq C, \forall i \in \{1,\dots,n\}, y^T \alpha = 0 \quad (22)$$

Where the Radius Basis function kernel function K (x, z) is used in this method to find the non-linear hyperplane in high dimension and such kernel function can be shown:

$$K(x,z) = e^{-\gamma ||x-z||^2} \quad (23)$$

The optimal non-linear hyperplane can be found by optimizing the (21) align to conditions (22).

IV. EXPERIMENT

A. Data labeling & Inter-Patient paradigm

1) MIT BIH Arrhythmia Dataset

Each record in MIT BIH Arrhythmia Dataset is split into 10s segments and labeled with binary label according to the conventional AAMI criteria which determine the type of beats during that 10s segment. We then split the patients randomly according to inter-patient paradigm into DS1 = {100, 101, 103, 106, 108, 109, 111, 112, 113, 114, 115, 116, 118, 119, 121, 122, 123, 124, 200, 201, 202, 203, 205, 207, 208, 209, 213, 219, 231} and DS2 = {105, 117, 214, 230, 232, 233, 234}, where DS1 : DS2 ~ 84 : 16 and 30% of DS1 will be used as validation during training, and DS2 will be kept uncovered for final evaluation. Recordings from patients 102, 104, 107, 217 are discarded because the beat annotations from those patients contain Q extensively so that they are too noisy to be used. The distribution of each dataset VS super classes is shown in table 1.

TABLE1. The distribution of MIT BIH Arrhythmia after segmentation

| Dataset | Number of segments in each super class | | | |
|---|---|---|---|---|
| | Normal | Abnormal | Noisy | Total |
| DS1 | 4156 | 2441 | 63 | 6660 |
| DS2 | 680 | 544 | 36 | 1260 |
| Total | 4836 | 2985 | 99 | 7920 |

2) MIT BIH Atrial Fibrillation Dataset

Each record in MIT BIH AFIB dataset is also split into 10s segments and labeled with binary label based on rhythm that the label locates in. Following the Inter-patient paradigm, patients are split into DS3 = {04015, 04043, 04048, 04126, 04908, 04936, 05091, 05261, 06426, 06995, 07162, 07859, 08405, 08434, 08455, 07910, 08215, 08219, 08378} and DS4 = {04746, 05121, 06453, 07879} where DS3: DS4 ~ 83 : 17 and 30% of DS3 will be used as validation during training, DS4 will be kept uncovered for final evaluation. The distribution of each dataset VS super classes is shown in table 2.

TABLE2. Inter-patient paradigm of MIT BIH Atrial Fibrillation dataset

| Dataset | Number of segments in each super class | | | |
|---|---|---|---|---|
| | AFIB | Non-AFIB | Noisy | Total |
| DS3 | 26843 | 42599 | 516 | 69958 |
| DS4 | 6502 | 7806 | 68 | 14376 |
| Total | 33345 | 50405 | 584 | 84334 |

B. Evaluation Metrics

In the experiment, we are going to use average precision score (AP score) to evaluate the binary classification performance of model and perform model selection. The AP score is equivalent to AUC PR but involves a better calculation measure the Precision and Recall information than simply approximate the area under the curve by trapezoidal rule. As shown in Table 1 and Table 2, the datasets we are working with are significantly imbalanced. Under such scenarios, researchers should be selective of evaluation metrics they adopted. Several typical evaluation metrics are used in the studies [11-15, 24-27] such as accuracy, specificity, sensitivity, F1, Area under the Receiver Operating Characteristic Curve (AUC ROC). However, specifically under a highly skewed imbalanced binary classes situation, singular evaluation metrics such as accuracy specificity sensitivity can be misleading since they are usually sensitive to distribution of dataset. ROC curves generally address those problems by reflecting the trade-offs between true positive and false positive [21]. However, in the study done by Fawcet et al. [20], shows that a curve dominant in ROC space if and only if its dominant Precision-Recall (PR) space and ROC curve usually overly optimistically evaluates the performance of a model under a highly skewed dataset situation. Conclusively speaking, it is inappropriate to use singular evaluation metrics when working with MIT BIH dataset and consequently using AUC PR is highly suggested as a more informative metric than AUC ROC.

The AP score is obtained by the average of precision at each threshold weighted by the amount of increase of Recall from last threshold:

$$AP\ score = \sum_{i}^{threshold} (Recall_i - Recall_{i-1})\ Precision_i \quad (24)$$

The Recall and Precision can be simply calculated as follows:

$$Precision = \frac{TP}{TP + FP} \quad (25)$$

$$Recall = \frac{TP}{TP + FN} \quad (26)$$

In order to have a comprehensive view of LSF model and comparison with existing model other metrics are also calculated including AUC ROC, accuracy, specificity. Where AUC ROC is the trade-off between true positive rate and false positive rate across different thresholds. Accuracy and specificity shown below:

$$Accuracy = \frac{TP + TN}{TP + FP + TN + FN} \quad (27)$$

$$Specificity = \frac{TN}{TN + FP} \quad (28)$$

C. Result

Table 3 presents a comparison of the LSF method and other existing methods trained only under intra-patient paradigm. The results show that our LSF method is comparable and even outperforms to many other methods in terms of evaluation metrics.

Similar to the study conducted by Minh et al. [18], our LSF model also experience a significant performance drop when transitioning from the intra-patient paradigm to the inter-patient paradigm. This performance gap arises due to the fact that when using the intra-patient paradigm, the test set contains data segments from the same patients as the training set, leading to data leakage. In order to assess the model's ability to generalize to unseen data and unseen patients, the inter-patient paradigm should be used. This paradigm is closer to realistic clinical applications, as it tests the model's performance on data from patients that were not included in the training set. By evaluating the model using the inter-patient paradigm, we can better understand its ability to handle new and unseen patients, which is crucial for real-world clinical applications.

Compared to the baseline model (LSTM with Fully connect layer) of our LSF method, our improved method not only shows a significant increase in AP score, but also generally performs better across other evaluation metrics. This improvement can be attributed to the combination of LSTM and SVM. The LSTM layer captures the time series properties and effectively separates the clusters of the two classes in high-dimensional space. The SVM classifier with RBF kernel can then easily track the hyperplane in this high-dimensional space. This combination is particularly effective when the dataset has a skewed distribution, as indicated by the AFIB detection test on DS4. Although there may be a drop in accuracy and recall when performing AFIB detection on DS4, several more comprehensive metrics, such as AP score and AUC ROC, show improvement. This further suggests that the SVM classifier enhances the generalization and robustness of the base model. These metrics provide a more accurate assessment of the model's performance and indicate its ability to handle real-world scenarios.

D. Ablation

In order to illustrate the efficacy of SVM block in our LSF method. The LSTM layers with a fully connected layer were trained using a specific loss function described in section III A. Two models were trained on DS1 and DS3 datasets, respectively. During the training process, 30% of the data was set aside as validation data. The data was shuffled after each epoch. The training process included an early stopping criterion. If the average precision (AP) validation score did not decrease over 10 consecutive epochs, the training would stop. The model with the highest validation score was recorded.

The performance of the base line model was tested on DS2 and DS3 datasets. The AP scores obtained from the testing are the performance of models be tested on DS2, DS3 respectively, showing that the AP score we got from DS2 is 0.9235 on DS2 and 0.9213 on, we got from DS3 is 0.9213, This score which is set as the base line to our LSF model. The plot of baseline test AP scores on DS2 and DS4 are shown in figure 4.

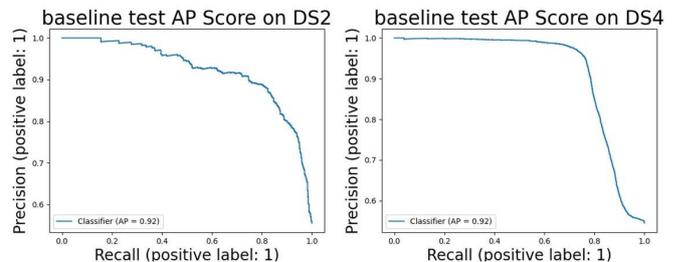

*Fig4. baseline test score of LSF model*

TABLE3. comparison between LSF and existing work follows intra-patient paradigm.

| Author/method | Intra or inter | Task | Signal duration | Accuracy | Recall | Specificity | AUC ROC | AP score |
|---|---|---|---|---|---|---|---|---|
| Cheng et al. [14] | intra | AFIB detection | 10s | 0.9752 | 0.9759 | 0.9740 | - | - |
| Zhang et al. [30] | intra | AFIB detection | Beat-wise | 0.9623 | 0.9592 | 0.9655 | - | - |
| Wang et al. [31] | intra | AFIB detection | 4s | 0.9740 | 0.9790 | 0.9710 | - | - |
| Xia et al. [32] | intra | AFIB detection | 5s | 0.9863 | 0.9879 | 0.9787 | - | - |
| LSF | Inter | AFIB detection | 10s | 0.8216 | 0.7029 | 0.9639 | 0.9442 | 0.9563 |
| LSF | Intra | AFIB detection | 10s | 0.9870 | 0.9891 | 0.9845 | 0.9984 | 0.9988 |

The hidden states of the last LSTM layer for DS1 and DS3 datasets were exported and used as input for two SVM classifiers. The SVM classifiers were trained on DS1 and DS3, respectively. Grid search was employed to find the optimal hyperparameters for the SVM classifiers. The hyperparameters considered were the penalty weight C, ranging from 0 to 2 with a step size of 0.1, and the class weight of each class, ranging from 0 to 1 with a step size of 0.1. The classifier with the best average precision (AP) score was selected based on the grid search results. This classifier was then integrated with the LSTM layer to form the LSF model. The LSF model was tested on DS2 and DS4 datasets, achieving AP scores of 0.9406 and 0.9563 respectively, shown in figure 5.

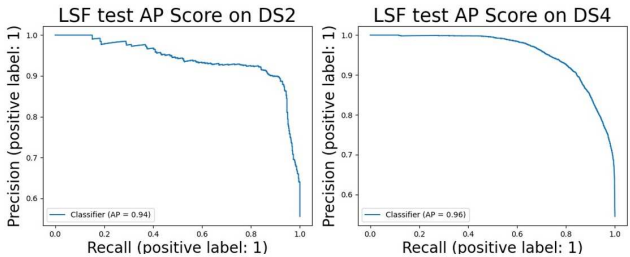

Fig5. LSF test score

Other evaluation metrics comparison between baseline and LSF on DS2 and DS3 shown below.

TABLE4. comparison of metrics performs arrhythmia detection tested on DS2.

| Arrhythmia detection test on DS2 | Accuracy | Recall | Specificity | AUC ROC | AP score |
|---|---|---|---|---|---|
| LSF | 0.8897 | 0.9029 | 0.8732 | 0.9333 | 0.9402 |
| baseline | 0.8341 | 0.8426 | 0.8235 | 0.9044 | 0.9235 |

TABLE5. comparison of metrics performs AFIB detection tested on DS4.

| AFIB detection test on DS4 | Accuracy | Recall | Specificity | AUC ROC | AP score |
|---|---|---|---|---|---|
| LSF | 0.8216 | 0.7029 | 0.9639 | 0.9442 | 0.9563 |
| Baseline | 0.8439 | 0.7747 | 0.9267 | 0.8672 | 0.9213 |

## V. CONCLUSION AND FUTURE WORK

Machine learning and deep learning methods have become increasingly popular in the healthcare field. However, it is crucial to exercise caution when applying these techniques. Inadequate use of these methods can result in biased models that struggle to generalize to future use cases. In this paper, we propose the LSF method to detect CVDs efficiently. Validated on the MIT-BIH arrhythmia and MIT-BIH AFIB datasets with the inter-patient paradigm used, LSF outperforms other related works in the way of efficiency and performance. In the future work, we aim to further test the real-time processing capability of our model and explore the possibility of deploying it on portable devices. Additionally, we plan to incorporate several other datasets to train the model, enhancing its generalization ability and expanding its applicability to a wider range of scenarios.